\documentstyle[aps,prb,graphicx,amsbsy,amssymb]{revtex}
\begin{document}

\twocolumn[\hsize\textwidth\columnwidth\hsize\csname
@twocolumnfalse\endcsname

\title{Guided Vortex Motion in Superconductors with a Square Antidot Array}
\author{A.V. Silhanek$^1$, L. Van Look$^1$, S. Raedts$^1$, R. Jonckheere$^2$, and V.V. Moshchalkov$^1$}

\address{$^1$Laboratorium voor Vaste-Stoffysica en Magnetisme,\\
K.U.Leuven, Celestijnenlaan 200D, B-3001 Leuven, Belgium.\\
$^2$Inter-University Micro-Electronics Center (Imec vzw),
Kapeldreef 75, B-3001 Leuven, Belgium}

\date{\today}

\maketitle

\begin{abstract}
We have measured the in-plane anisotropy of the vortex mobility in a thin Pb film with a square array of antidots. The Lorentz force, acting on the vortices, was rotated by adding two perpendicular currents and keeping the amplitude of the net current constant. One set of voltage probes was used to detect the vortex motion. We show that the pinning landscape provided by the square antidot lattice influences the vortex motion in two different ways. First, the modulus of the vortex velocity becomes angular dependent with a lower mobility along the diagonals of the pinning array. Second, the vortex displacement is preferentially parallel to the principal axes of the underlying pinning lattice, giving rise to a misalignment between the vortex velocity and the applied Lorentz force. We show that this anisotropic vortex motion is temperature dependent and progressively fades out when approaching the normal state.
\end{abstract}

\vskip1pc] \narrowtext


 The interaction of a moving elastic medium
with an array of obstacles has received much attention during the
last years, in part due to a variety of condensed matter
systems that can be described within this
model.\cite{martinoli78prb,runge93euro,vanblaaderen97nature,hu97prb}
In general, the symmetry of the substrate over which the elastic
medium moves, should also be reflected in the dynamic properties
of the system.

Particular interest has gone to the study of the flux line lattice
(FLL) in superconductors with an artificial pinning potential,
such as randomly distributed columnar tracks,\cite{suppression} arrays of
submicron holes (``{\it
antidots}")\cite{baert95prl,vvm96prb,vvm98prb} or magnetic
dots.\cite{morgan98prl,vanbael99prb}

In a superconductor with a random distribution of pinning centers,
an isotropic vortex mobility is expected with a vortex motion parallel
to the Lorentz force. On the other hand, if the
rotational symmetry of the pinning potential is broken, the
response will be anisotropic,\cite{vanlook02prb} and guided vortex
motion may appear.

Recent experiments\cite{pastoriza} have shown that the presence of
twin boundaries in high temperature superconductors leads to a
preferential motion of the FLL along the twinning planes. Guided
vortex motion has also been found in systems with a four-fold
symmetry, like square arrays of Josephson junctions.\cite{marconicandia} Theoretical work on vortex motion in a
square array of pinning centers has forecasted guided vortex
motion or \textit{channeling} along the principal axes of the
array, and along its diagonals.\cite{clecio,devilstaircases,carneiro,marconi} In
these systems, the vortex velocity $\bf v$ preferentially snaps to
the symmetry axes of the pinning array, leading to a misalignment
between vortex velocity and the applied driving force $\bf
f$. Even though this
lock-in transition has been predicted in many numerical simulations
for a square array of pinning centers, it had not yet been observed
experimentally.

In this work, we report on electro-transport measurements on a Pb
thin film with a square array of antidots, for the whole $360^\circ$ range
of in-plane current orientations. This allows us to determine the
misalignment between the applied current and the electric field
generated by the vortex movement. We show clear evidence of guided
vortex motion along the principal axes of the antidot array, and
determine the trapping angle in the studied structure.


The used sample is a 50-nm-thick Pb film, with a square antidot lattice (antidot size $b^2=0.6 \times 0.6 \mu m^2$) of period $d=1.5 \mu m$, which corresponds to the first matching field $H_1=\Phi_0/d^2=$9.2 G, with $\Phi_0$ the superconducting flux quantum. We patterned the Pb film in a cross-shaped geometry (see Fig. \ref{fig1}) to allow electrical
transport measurements at any in-plane orientation of the applied current. The cross consists of two 300~$\mu$m wide strips containing the square array of antidots aligned with the principal axes of the cross.

\begin{figure}[htb]
\centering
\includegraphics[angle=0,width=90mm]{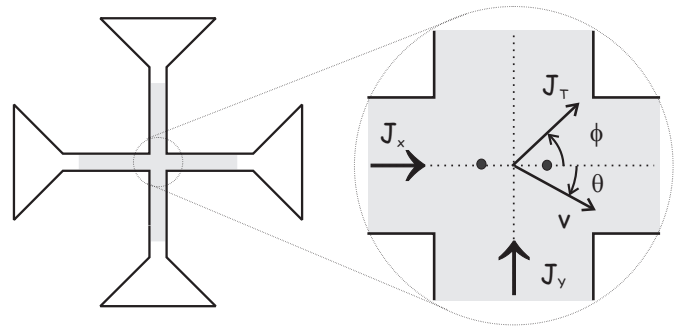}
\caption[]{{\small Left: sketch of the cross-shaped Pb film with a square antidot array (grey area). Right: zoom in of the central part of the cross and orientation of the total applied current ${\bf J_T}$ and vortex velocity $\bf v$ with respect to the $\hat {\bf x}$-axis of the pinning array. The black dots represent the voltage probes.}}
\label{fig1}
\end{figure}

This pattern was prepared by electron-beam lithography in a polymethyl metacrylate/methyl metacrylate (PMMA/MMA) resist bilayer covering the SiO$_2$ substrate. A Ge(20~\AA)/Pb(500~\AA)/Ge(200~\AA) film was then electron-beam evaporated onto this mask while keeping the substrate at 77 K. Finally, the resist was removed in a liftoff procedure in warm acetone. The sample has a T$_c$=7.220 K with a transition width of 7 mK for $H$=0.

The transport measurements were performed in a $^3$He cryostat using two independent dc currents, $J_X$ and $J_Y$, applied at the extremes of the cross legs, allowing to control the orientation ($\phi$) of the total current ${\bf J_T} = J_X \cos \phi ~{\hat {\bf x}}+J_Y \sin \phi ~{\hat {\bf y}}$, with respect to the principal axes of the square pinning array. The voltage contacts were aligned with the $\hat {\bf x}$ direction (see Fig.~\ref{fig1}, right panel). In the $V_X(\phi)$ measurements, we sweep the angle $\phi$ of the total current ${\bf J_T}$ while keeping its amplitude $J_T=\sqrt{J_X^2+J_Y^2}$ constant. The magnetic field $\bf H$ was applied perpendicular to the film surface.

It is worth to notice that in previous works,\cite{pastoriza,loussane} in order to determine the direction of the vortex movement, two sets of voltage probes arranged perpendicularly (aligned with each of the current directions), have been used. The simultaneous acquisition of the independent components of the electrical field ${\bf E} = {\bf v} \times \bf H$, allows one to determine the modulus and direction of the average vortex velocity $\bf v$ and therefore the misalignment between $\bf v$ and the applied Lorentz force ${\bf f} = {\bf J_T} \times {\bf \Phi_0} $. Although this procedure is strictly necessary for uniaxial systems, simple symmetry considerations associated with the periodic pinning landscape show that for higher degree of symmetry, it is possible to obtain the same information with a {\it single pair} of voltage contacts. Indeed, in our experiment, the intrinsic four-fold symmetry imposed by the pinning potential implies that $V_Y(\phi)=V_X(\phi+90^\circ)$, and therefore by measuring $V_X(\phi)$ in the whole angular range, we can deduce $V_Y(\phi)$.

\begin{figure}[htb]
\centering
\includegraphics[angle=0,width=90mm]{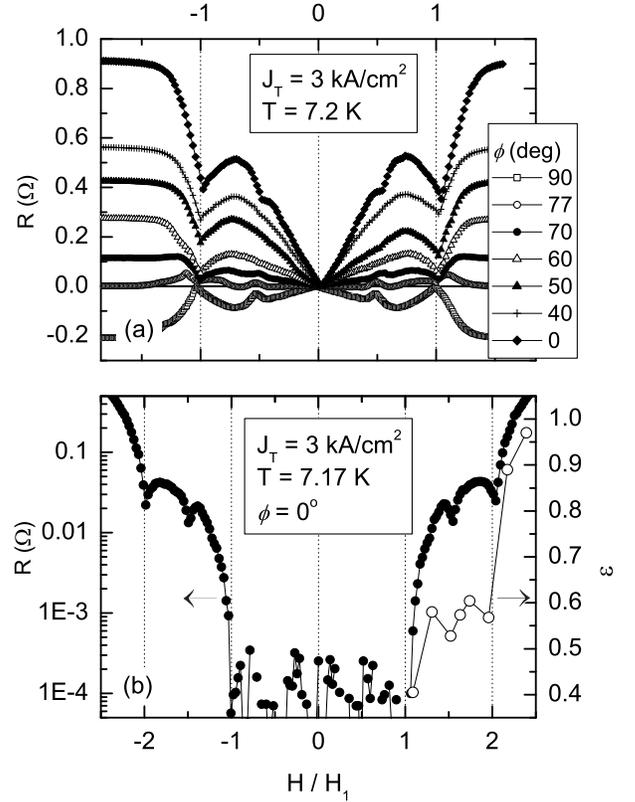}
\caption[]{{\small Magnetoresistance $R(H)$ at $J_T$= $3 kA/cm^2$ for (a) several angles $\phi$ of the total current ${\bf J_T}$ at $T=7.20$ K (b) $\phi=0^\circ$ and $T=7.17$ K. In (b), the field dependence of the eccentricity factor $\varepsilon=V(\phi=45^\circ)/V(\phi=0^\circ)$ is shown (open symbols).}}
\label{fig2}
\end{figure}

In Fig.~\ref{fig2}(a) we show the field dependence of the resistance $R(H)$ for $J_T$= $3 kA/cm^2$ and $T=$7.20 K, for several angles $\phi$ of the total current. The most obvious features of this figure are the well defined dips at $1/2 H_{1}$ and $H_1$, as a consequence of the reduction of the FLL mobility, reminiscence of the vortex configurations energetically stable in the equilibrium state. For low vortex density ($H<0.4H_1$), a linear field dependence characteristic of a flux flow regime $\rho_{FF}=\rho_n H/H_{c2}$, with $\rho_n$ the normal state resistivity and $H_{c2}$ the upper critical field, is observed.\cite{rosseel}

It should be noted that, since the experiments were carried out very close to $T_c$, temperature stability plays a crucial role. For instance, the curve at $\phi=50^\circ$ in Fig.~\ref{fig2}(a) exhibits a small asymmetry between the positive and negative branches of the magnetoresistance ($R(H) \neq R(-H)$) because of a temperature drift of only $0.2$mK.

Due to the misalignment between the wire bounded voltage probes and the lithographically defined current contacts, the resistance in the normal state reaches the condition $R=0$ (i.e. $V_X=0$)  at $\phi=77^\circ$ rather than at $\phi=90^\circ$. Remarkably, a finite resistance appears when penetrating in the superconducting phase, $\left| H \right| \leq 15$ G $\approx 1.6 H_1$ (open cicrcles in Fig. \ref{fig2}(a)).
Since the $R(H)$ curve at $\phi=77^\circ$ is symmetric around $H=0$, we rule out the possibility to attribute this dissipation to the Hall effect. We have also checked that thermoelectric voltages, generated by thermal gradients, have a negligible contribution to the total voltage. This ``even Hall effect"
might be attributed to guided vortex motion,\cite{pastoriza,loussane} however we have observed that this effect persists in plain films {\it without any periodic structure}, where transport properties should be isotropic. Similar anomalies  have also been reported in the transverse voltage in Josephson Junction arrays\cite{mooij} and in the longitudinal voltage in plain films.\cite{kwong} Although there is no general consensus about the origin of this effect, a possible explanation involves proximity effects due to the N/S boundary in the voltage contacts.\cite{park}

\begin{figure}[htb]
\centering
\includegraphics[angle=0,width=70mm]{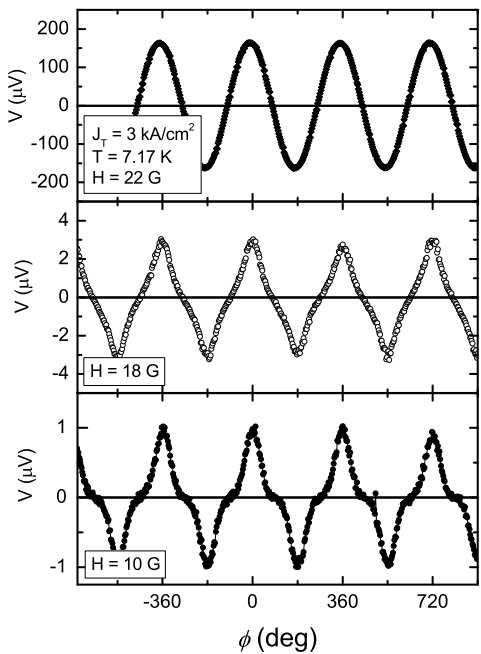}
\caption[]{{\small Angular dependence of the voltage $V_X(\phi)$ for $J_T=3 kA/cm^2$ and $T=$7.17 K, at $H=$10, 18 and 22 G.}}
\label{fig3}
\end{figure}

Fig. \ref{fig2}(b) shows $R(H)$ for $\phi=0^\circ$ and $J_T=3 kA/cm^2$ at $T=$7.17 K. For fields $\left| H \right| < H_1$, where vortices are strongly pinned at the antidots, the dissipation lies below our experimental resolution. At this temperature, the maximum number $n_s$ of flux quanta that an antidot can hold is $n_s \approx b/4\xi(T) \sim 0.3 \ll 1$,\cite{schmidt} where $\xi(T)$ is the superconducting coherence length. Under these circumstances, interstitial vortices appear in the sample for magnetic fields  $H >H_1$. They are weakly pinned (``caged") at the interstitial positions between the repulsive saturated antidots. The matching features at  $\frac{3}{2}H_1$ and $2H_1$ are a result of their stable vortex configurations. In this scenario, where two species of vortices coexist, the motion of interstitial vortices should be strongly affected by the presence of those vortices which remain pinned. As field progressively increases, vortex-vortex interaction becomes more relevant in comparison with the vortex-pinning energy and eventually, close to the upper critical field $H_{c2}$, a vortex liquid phase appears. In order to study the role of the periodic pinning array on the vortex dynamics in each of these regimes, we have measured the in-plane angular dependence of the voltage $V_X(\phi)$ for the same conditions of Fig. \ref{fig2}(b) ($J_T=3 kA/cm^2$ and $T=$7.17 K) at $H=$10, 12, 14, 15, 16, 18, 20 and 22 G. The curves $V_X(\phi)$ for some of them are presented in Fig. \ref{fig3}.


The upper panel of Fig. \ref{fig3} shows $V_X(\phi)$ close to the normal state ($H_{c2}(7.17 K) \approx 25$ G). The recorded signal follows a smooth sine function dependency characteristic of an isotropic regime with a constant vortex velocity independent of the current orientation. At $H=$18 G, near to the second matching field, the response is quite different. Now, a sawtooth like curve clearly manifests the existence of an in-plane anisotropy in the vortex movement. This anisotropy becomes even more evident as $H$ approaches to the first matching condition (lower panel). As we will show below, the characteristic shape of $V_X(\phi)$, alternating well pronounced peaks and plateaus, is a fingerprint of guided vortex motion.

\begin{figure}[htb]
\centering
\includegraphics[angle=0,width=90mm]{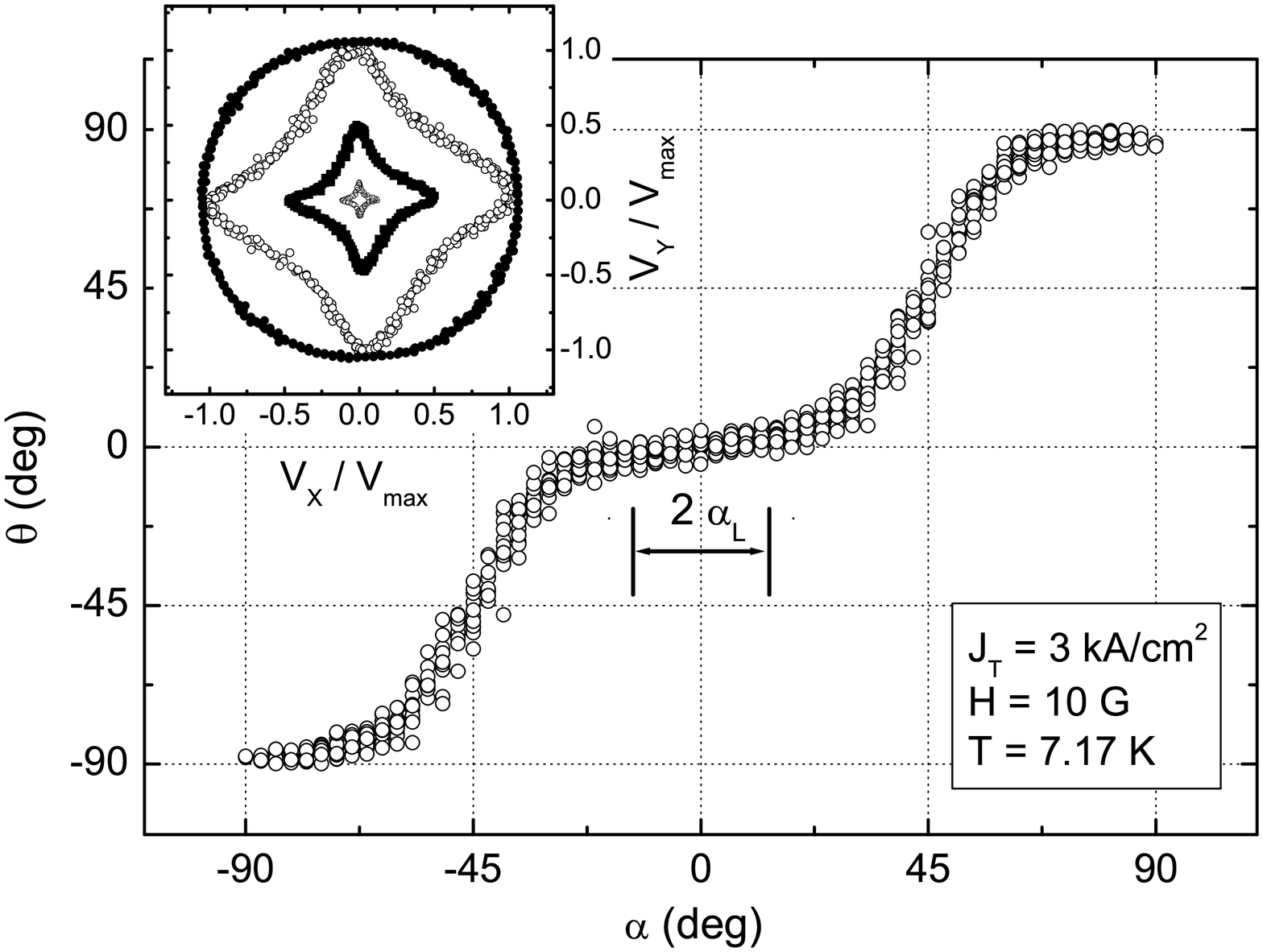}
\caption[]{{\small Angle $\theta$ of the vortex velocity as a function of the Lorentz force orientation $\alpha$ for $J_T=3 kA/cm^2$ and $T=$7.17 K, at $H=$10 G. Inset: polar graph $V_X \times V_Y$ for $J_T=3 kA/cm^2$ and $T=$7.17 K, at (from inner to outer) $H=$10, 14, 16 and 22 G.}}
\label{fig4}
\end{figure}

As we pointed out previously, knowing the longitudinal component of the voltage $V_X(\phi)$, we can deduce the transversal component $V_Y(\phi)=V_X(\phi+90^\circ)$. This allows us to build a polar graph $V_X \times V_Y$, as shown in the inset of Fig. \ref{fig4} for $H=$10, 14, 16 and 22 G, where the curves have been scaled for the sake of clarity. The circular shape of the outer curve ($H=$22 G) reflects the rotational invariance of an isotropic conductor. As field decreases, the shape of the polar plot smoothly evolves from a circle, through a rhombus (not shown) and eventually data lies along an {\it astroid} for $\left| H \right| <$ 20 G, indicating that vortices move faster along the principal axes $[\pm1,0],[0,\pm1]$ (easy direction) of the pinning structure than along the diagonals (hard direction).

The degree of anisotropy of the vortex velocity is given by the eccentricity factor $\varepsilon=V(\phi=45^\circ)/V(\phi=0^\circ)$. In case of a circular shape, $\varepsilon=1$, whereas for a rhombus $\varepsilon=1/\sqrt 2$. In Fig. \ref{fig2}(b) we show the anisotropy $\varepsilon$ from the analysis of the data at T=7.17 K,  as a function of field. We clearly observe that $\varepsilon(H)$ reproduces the behavior of the magnetoresistance $R(H)$. According to the theory,\cite{marconicandia} the vortex dynamics should become more anisotropic as $J_T$ decreases towards the depinning current $J_{dp}$. On the other hand, at constant $J_T$, a higher $J_{dp}$ yields a lower dissipation. Therefore, a stronger anisotropy should be detected where $R(H)$ has local minima, in agreement with our observations. A similar $\varepsilon(H)$ behavior, but somewhat less pronounced was found at higher temperatures for all fields $\left| H \right| < H_{c2}$. This result indicates that the anisotropic vortex motion is present regardless of the dynamic regime.

Once the polar graph is built, it is possible to extract the angle of the electric field $\theta_E=Atan(V_Y/V_X)$ and hence the angle of the vortex velocity $\theta=\theta_E+90^\circ$ as a function of the Lorentz force orientation $\alpha=\phi-90^\circ$. In the main panel of Fig. \ref{fig4} we show $\theta(\alpha)$ so obtained for $H=$10 G, folded to the fourth and first quadrant. For an isotropic conductor, $\theta=\alpha$ and a straight line should be observed. In contrast, a highly non-linear response is found.

Starting from $\alpha=0$ and rotating the current in counter-clockwise direction, i.e. increasing the angle $\phi$, we observe that for low values of $\alpha$, the force $\bf f$ lies ahead of $\bf v$, which remains close to the $[1,0]$ axis ($\theta \sim 0^\circ$). The misalignment $\left |\alpha-\theta\right |$ reaches its maximum value at the lock-in angle $\alpha_L \sim 15^\circ$, beyond which the vortex movement is no longer restricted to follow the $[1,0]$ direction and a fast increase in $\theta$ is observed. At the $[1,1]$ orientation, force and velocity are aligned due to the four-fold symmetry of the underlying pinning potential. Above $[1,1]$, the Lorentz force lags behind the velocity vector which rapidly approaches to the $[0,1]$ axis. This is the central result of the paper which clearly indicates the role played by the periodic pinning potential in the vortex dynamics.


Recent simulations\cite{clecio} for a moving FLL in a superconducting films with square pinning array show that minor lock-in phases should also be present at $\pm 45^\circ$ orientations. Further features might be present near the driving angles along the symmetry directions of the pinning array other than the principal axes.\cite{devilstaircases} However, no evidence of such phases were found in our experiments. The absence of these regimes may be attributed to the relatively large size of the antidots. Indeed, since the pinning potential is averaged in the direction of motion, it is expected that the movement at $\pm 45^\circ$ (diagonals of the square antidots) be progressively washed out as the size of the antidots increases. Well defined channels will exist only if $d/4b > 1$. In our experimental conditions $d/4b=0.625$ and thus channeling effects should be suppressed. This assumption is confirmed in the weakly coupled superconducting wire network limit by simulations for square Josephson Junction arrays by Marconi et al.\cite{marconi} where no particular effects appear at $\pm 45^\circ$.


In summary, using dc-transport measurements we have demonstrated that the interaction of a  moving flux line lattice with a periodic pinning potential, breaks the rotational symmetry and induces a guided vortex motion along the principal symmetry orientations of the pinning array. As a consequence, we found that for orientations different than the symmetry directions, the vortex velocity is not aligned with the Lorentz force. This effect vanishes approaching to the normal state as the vortex movement becomes insensitive to the pinning symmetry.

We would like to thank Clecio C. de Souza Silva for helpful discussions. This work was supported by the ESF ``Vortex" Program, the Belgian Interuniversity Attraction Poles (IUAP), the Flemish GOA and FWO Programs.

\bibliographystyle{prsty}

\end{document}